# SVPLEX: A Nextflow Pipeline for Cohort-level Structural Variant Calling

Jacob E. Munro[1,2], Mark F. Bennett[1,2,3], Melanie Bahlo[1,2]

1. Walter and Eliza Hall Institute of Medical Research, Parkville, Victoria, Australia
2. Department of Medical Biology, University of Melbourne, Parkville, Victoria, Australia
3. Epilepsy Research Centre, Department of Medicine (Austin Health), University of Melbourne, Heidelberg, Victoria, Australia

## Summary

**SVPLEX** is a Nextflow pipeline for cohort-level structural variant detection from short-read whole-genome sequencing data. The pipeline implements six different structural variant callers with different strengths and weaknesses, integrating different levels of evidence for SVs, and generates a merged consensus callset across the analysis cohort. Callset filtering is achieved by leveraging consensus among multiple individual callers and by ensuring that deletion and duplication calls are supported by observable changes in read depth. The output merged cohort SV callset can then be used to assess cohort-specific variation, remove technical artefacts, and serve as input for rare disease variant prioritisation workflows.

**SVPLEX** is user-friendly, reproducible, scalable, and can be executed flexibly on either a local workstation, a high-performance compute (HPC) cluster, or deployed on cloud infrastructure. The required inputs are alignment files for the cohort of interest, and the output is a single merged cohort structural variant VCF. **SVPLEX** is available on GitHub (bahlolab/SVPLEX) and is licensed under the MIT open-source licence.

## Statement of need

Structural variants (SVs) encompass a range of DNA alterations larger than 50 bp[1], including deletions, duplications, insertions, inversions and translocations, as well as complex events comprised of multiple concurrent simple events. Due to the higher complexity, detecting SVs from short-read whole-genome sequencing data is more challenging than detecting short variants (single nucleotide variants and indels). Many methods have been developed for SV calling, using different lines of evidence from the sequencing data and assessing different SV classes[2]. Unifying evidence from multiple SV callers across individuals in a cohort is difficult, with few pipelines in existence that can harmonise SV calls. In contrast, germline short variant calling has converged on a smaller number of widely adopted methods with similar levels of performance[3]. It is not standard practice to employ multiple short variant callers outside of certain niche cases (e.g., somatic mosaic variant detection[4]).

SVs are detected based on a range of evidence types, including: 1) read-pair (RP) evidence, which detects aberrations in the expected orientation and spacing of paired-end reads, 2) split-read (SR) evidence, which detects SV breakpoints wherein a portion of a sequencing read maps to a given locus in the reference genome and another portion of the read maps to a distal locus or not at all, and 3) read-depth (RD) evidence, which detects variation in read-depth from background that underlie deletions and duplications[5,6]. Individual callers make use of different pieces of evidence; SR and RP evidence are typically interrogated together, and are employed by tools such as Delly, LUMPY, Manta and Dysgu to detect deletions, duplications, insertions, inversions and translocations[7–10]. In contrast, RD-based callers such as Delly-CNV and CNVnator rely on read-depth segmentation to identify copy number variants[7,11]. RD-based callers provide complementary evidence to SR/RP callers and can detect CNVs without mappable breakpoint-spanning reads, although breakpoint resolution is poorer and sensitivity is reduced for smaller events, particularly those below one kilobase in length[1].

Because no single caller outperforms all others under all conditions or for all variant types, it is standard practice to run multiple SV callers[12,13]. A straightforward approach is to run a handful of well-performing tools and take either the union of calls if sensitivity is a priority, or the intersection if specificity is a greater concern. More advanced approaches use ensemble methods, wherein machine learning is used to combine evidence from multiple individual callers to generate more refined predictions[14–16].

An ideal pipeline for rare disease cohort analysis will integrate the outputs from multiple callers, perform consensus filtering, and merge calls into a unified cohort callset. This enables two key aspects of variant prioritisation in rare disease cohorts: 1) consistent application of inheritance model-based filtering strategies for familial data, and 2) identification and removal of cohort or batch-level artefacts. De novo variants, including SVs, are a major cause of Mendelian diseases[17]. However, effective de novo filtering requires that SV calls are unified between parents and probands, as otherwise calls with even slightly differing breakpoint representations can be incorrectly classified as de novo.

SV callsets are sensitive to both the underlying sequencing data (e.g. platform, library preparation, coverage), as well as the software used to call them. A unified cohort thus provides an effective way for cohort-level allele frequency filtering to be applied, removing common SV calls, whether real or artefactual, from downstream prioritisation. Filtering based on frequency in reference databases such as gnomAD is insufficient on its own, as the SV databases are dependent on both the underlying data and data processing used[18], and may not transfer directly to a cohort processed with a different pipeline.

## State of the field

Ensemble calling and cohort-level merging present several challenges. Firstly, although most SV callers produce VCF files, they don't necessarily use the same conventions for variant representation. Second, due to the differing evidence used, callers rarely report identical breakpoints for the same underlying event. Thus, prior to any sample-level or cohort-level merging, call representations must be harmonised and an approximate matching of equivalent events performed. Several tools have been developed for this purpose, including Truvari, SVDB, and SURVIVOR, which match SVs using user-configurable similarity criteria including breakpoint distance, reciprocal overlap, sequence similarity, and size similarity[19–21].

Individual SV callers have been shown to have gaps in detection, with ensemble callers typically outperforming individual tools[14–16]. One such tool is Parliament2[15], which combines results from up to six individual callers (five based on SR/RP evidence, one on RD evidence), merges their outputs with SURVIVOR, and assigns a rule-based variant confidence score based on the supporting caller combination and SV size. Confidence scores are derived from the HG002 SV benchmark dataset[22], which is limited to deletions and insertions, making it unclear how well the derived scoring rules generalise to other SV classes or sequencing datasets. A further limitation of Parliament2 and related ensemble callers is that they generate per-sample consensus callsets rather than performing cohort-level merging.

Few workflows implement both multi-caller discovery and cohort-level merging. GATK-SV is one such cohort-scale pipeline that integrates four SR/RP callers and two RD-based callers, and performs harmonisation and additional filtering[23]. However, GATK-SV is built specifically for the Google Cloud Platform; it does not run on other cloud platforms or HPC infrastructure, limiting access for groups whose data is restricted to a specific platform or to on-premises compute.

**SVPLEX** provides an alternative framework for cohort-level SV detection and can be readily deployed across the wide range of execution environments supported by Nextflow, including institutional HPCs. It implements six short-read SV callers: four based on SR/RP evidence and two on RD evidence. Including two RD callers allows large CNVs lacking SR/RP evidence to receive independent support from more than one caller. **SVPLEX** does not rely on benchmark-derived confidence scores or trained models, but instead applies transparent, user-configurable filtering based on 1) the number of callers supporting an individual event, and 2) additional read-depth support, as quantified by Duphold[24].

## Software design

**SVPLEX** is implemented in Nextflow, providing reproducibility, containerisation, scalability and execution on multiple compute backends[25]. Parallelism and software isolation are achieved by running each task individually by sample and caller. Software dependencies are provided primarily by Biocontainers[26], with Seqera Wave containers used where multi-tool containers are required. Conda recipes using Bioconda[27] packages are provided as a fallback in cases where container engines such as Apptainer are unavailable. The target audience for **SVPLEX** is researchers analysing human short-read whole-genome sequencing data. The pipeline is fully compatible with the Seqera Platform, enabling graphical launch, monitoring, and execution reporting for users who prefer not to use the command line.

**SVPLEX** implements four callers using SR/RP evidence, Manta, Dysgu, Smoove (a wrapper for LUMPY), and Delly, as well as two RD callers, CNVnator and Delly-CNV (Figure 1). These tools were selected due to their maturity and widespread usage. The user may run only a subset of callers for faster computation; however, at least two SR/RP and two RD callers are recommended for consensus filtering. Each tool is implemented in an independent sub-workflow, with caller-specific post-processing applied to standardise the outputs. For example, CNVnator calls are converted from a generic tabular format to VCF files, and Delly-CNV calls are normalised such that they are represented as either deletions (SVTYPE=DEL) or duplications (SVTYPE=DUP) rather than generic CNVs (SVTYPE=CNV).

Users must provide a list of BAM or CRAM format alignment files, a reference genome FASTA file, and optionally a pedigree file for family-based calling. Auxiliary reference files required by certain tools are fetched automatically for supported GRCh37 and GRCh38 references. A test profile and test-fixture generation script are included to support testing and validation of the pipeline in new execution environments.

An optional familial mode enables joint genotyping of related samples to improve call consistency and support inheritance-based variant prioritisation. Joint analysis of families can rescue low-evidence SV calls that may otherwise fail per-sample thresholds and improve the consistency of breakpoint representation within families. This can reduce Mendelian inconsistencies that may cause pathogenic variants to be missed or variants to be incorrectly classified as de novo[28]. This mode is only supported by the Delly, Delly-CNV, Smoove and Manta SV callers.

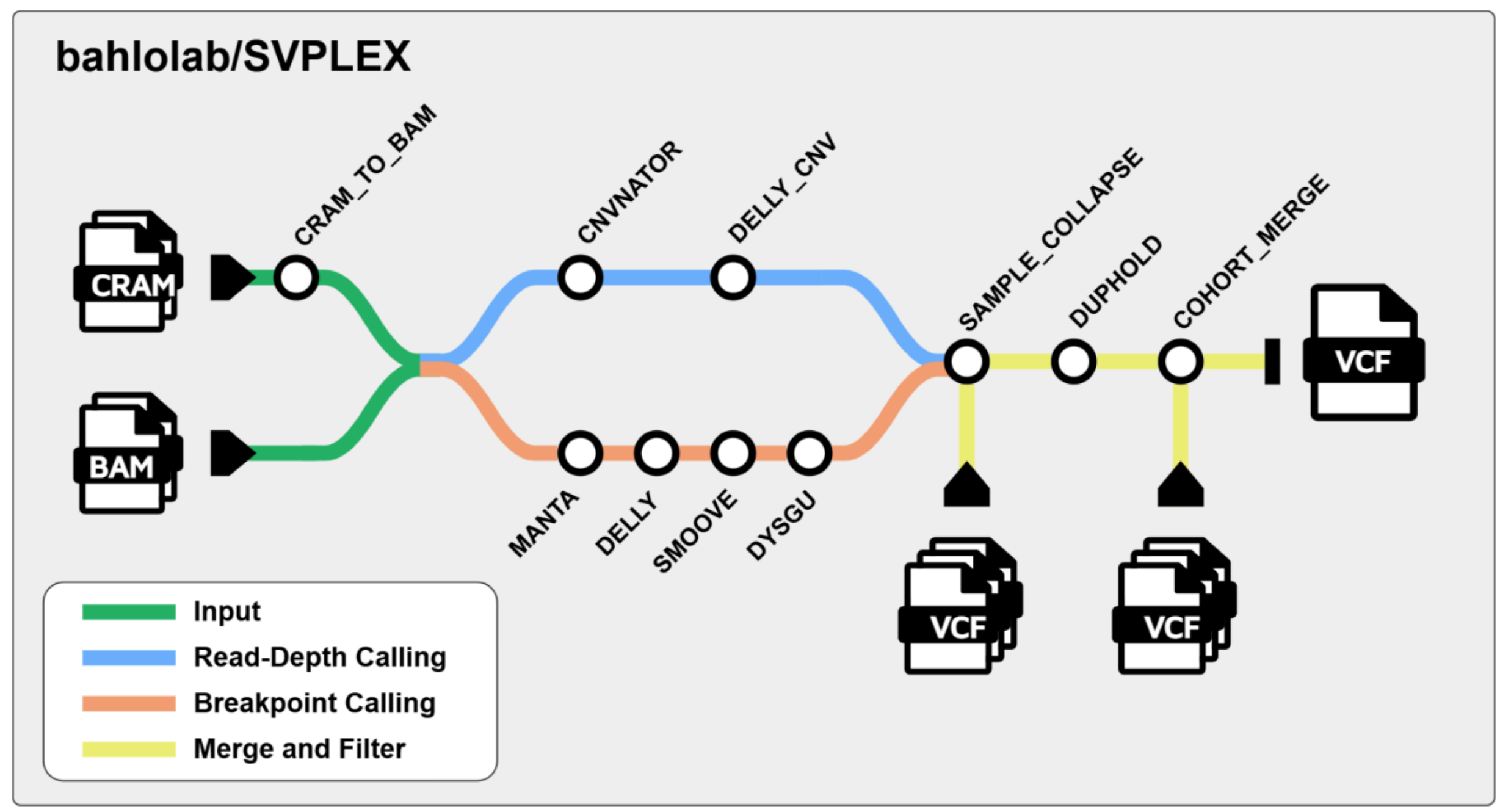

***Figure 1**: Schematic overview of the **SVPLEX** pipeline.*

**SVPLEX** implements three interchangeable SV merging backends: SVDB, Truvari, and Matcha. Each exposes different merging parameters and may suit different use cases; SVDB is chosen as the default. Merging is performed once for each sample, collapsing calls derived from multiple callers, then a final time merging all samples into a single cohort VCF. For SVDB and Truvari, merging is performed separately for interval-like events (deletions, duplications, and inversions) and for breakend-like events (translocations, insertions, unresolved breakends). This allows individual merging parameters to be configured separately for different variant classes, with interval-like events relying more on reciprocal overlap than breakend proximity by default.

Variant filtering is user configurable. By default, a variant must be supported by at least two callers to be retained in a sample-level callset. Deletions and duplications may be further filtered using Duphold[24], which quantifies evidence of decreased or increased read-depth for deletions and duplications, respectively.

**SVPLEX** also provides multiple entry points into the pipeline, so that previously completed, computationally intensive processes can be reused. Caller-specific single-sample VCFs may be provided to skip variant calling and proceed directly to sample-level collapsing. Alternatively, merged and filtered single-sample VCFs may be provided to skip both calling and sample-level collapsing and proceed directly to cohort-level merging. This is particularly useful for combining cohorts processed in separate batches.

## Research impact statement

**SVPLEX** has been used to search for pathogenic structural variants in a large cohort of individuals with developmental and epileptic encephalopathy (DEE), with 25% of diagnoses in the cohort being attributed to structural variants[29]. Additionally, **SVPLEX** was used to identify a complex structural variant in *FBRSL1* associated with a severe DEE[30]. **SVPLEX** is also used to support ongoing analysis of disease cohorts within the Bahlo Laboratory at the Walter and Eliza Hall Institute.

## AI usage disclosure

**SVPLEX** was originally written without generative AI assistance. Since May 7, 2026 (GitHub commit '0b23994'), Claude Code has been used for coding and documentation assistance, using a mixture of the Sonnet 4.6 and Opus 4.7 models depending on task complexity. The AI tools were given clear and specific instructions on implementation and design, and all AI-generated content has been reviewed and tested by the authors. The authors made all key decisions and take full responsibility for the reliability and maintenance of the software. The Claude web interface has been used to assist with grammar, clarity, and phrasing for this manuscript.

## Acknowledgements

This research was supported by the Commonwealth through an Australian Government Research Training Program Scholarship (DOI: https://doi.org/10.82133/C42F-K220).

This work was made possible through the Victorian State Government Operational Infrastructure Support Program and the Australian Government NHMRC IRIISS.

MB was funded by an Australian National Health and Medical Research Council Investigator grant (APP1195236). MFB was supported by an MRFF Genomics Health Futures Mission Grant (2007707). JEM was supported by a WEHI-CSL translational data science PhD top-up scholarship.